\documentclass[twocolumn]{aastex62}

\usepackage{epsfig}
\usepackage{epstopdf}
\usepackage{graphics}
\usepackage{xcolor}

\usepackage{array}
\usepackage{booktabs}
\usepackage{amsmath,amssymb,amsfonts,amsbsy}
\usepackage{mathrsfs}
\usepackage{url}
\usepackage{multirow}

\usepackage{color}
\usepackage{ulem}
\usepackage{enumerate}

\usepackage{lineno}
\renewcommand{\figurename}{Fig.}
\renewcommand{\tablename}{Tab.}
\begin{document}

\title{\bf {Constraining the position of the knee in the galactic cosmic ray spectrum with ultra-high-energy diffuse $\gamma$-rays}}

\correspondingauthor{Yi-Qing Guo, Bing-Qiang Qiao and Wei Liu}
\email{guoyq@ihep.ac.cn; qiaobq@ihep.ac.cn; liuwei@ihep.ac.cn}

\author{Pei-Pei Zhang}
\affiliation{Key Laboratory of Dark Matter and Space Astronomy, Purple Mountain Observatory,
Chinese Academy of Sciences, Nanjing 210023, China}
\affiliation{School of Astronomy and Space Science, University of Science and Technology of China, Hefei 230026, China}
\affiliation{Key Laboratory of Particle Astrophysics, Institute of High Energy Physics, Chinese Academy of Sciences, Beijing 100049, China}

\author{Yi-Qing Guo}
\affiliation{Key Laboratory of Particle Astrophysics, Institute of High Energy Physics, Chinese Academy of Sciences, Beijing 100049, China}
\affiliation{University of Chinese Academy of Sciences, Beijing 100049, China}

\author{Bing-Qiang Qiao}
\affiliation{Key Laboratory of Particle Astrophysics, Institute of High Energy Physics, Chinese Academy of Sciences, Beijing 100049, China}

\author{Wei Liu}
\affiliation{Key Laboratory of Particle Astrophysics, Institute of High Energy Physics, Chinese Academy of Sciences, Beijing 100049, China}

\begin{abstract}
{The diffuse $\gamma$-ray emission was measured up to $957$ TeV by the Tibet-AS$\gamma$ experiment recently. Assuming that it is produced by the hadronic interaction between cosmic ray nuclei and the interstellar medium, it requires that the cosmic ray nuclei should be accelerated well beyond PeV energies. Measurements of the cosmic ray spectra for different species show diverse results at present. The Tibet experiments showed that the spectrum of proton plus helium has an early knee below PeV. If this is correct, the diffuse $\gamma$-ray emission would suggest an additional component of Galactic cosmic rays above PeV energies. This second component may originate from a source population of so-called PeVatrons revealed by recent ultra-high energy $\gamma$-ray observations, and could contribute to the cosmic ray fluxes up to the energy of the second knee. On the other hand, the KASCADE measurement showed that the knee of protons is higher than PeV. In this case, the diffuse $\gamma$-rays observed by Tibet-AS$\gamma$ can be well accounted for by only one cosmic ray component. These two scenarious (i.e. the Tibet and KASCADE knees) could be distinguished by the spectral structures of diffuse $\gamma$-rays and cosmic ray nuclei. Future measurements of spectra of individual nuclei by HERD and LHAASO experiments and diffuse $\gamma$-rays by LHAASO can jointly constrain these two scenarios.}
\end{abstract}

\section{Introduction}
\label{sec:intro}
It is generally believed that the knee of cosmic rays (CR) \citep{1958On} is due to the acceleration limit of Galactic CR sources \citep{1961NCim...22..800P,1993A&A...274..902S,1997JPhG...23..979E,2002PhRvD..66h3004K,2007ApJ...661L.175B}. The individual nuclear spectrum plays a very important role to unveil its origin. Due to the limited composition resolution for ground-based experiments, it is challenging to precisely measure the CR composition at the knee region. Some experiments have devoted their efforts to carry out the spectral measurements of individual species, but with very large uncertainties. 
Basically there are two different results on the spectra of light components in CRs around the knee.
The results from KASCADE experiment showed that the knee-like structure of protons is higher than PeV \citep{2005APh....24....1A}. However, the AS$\gamma$-YBJ, ARGO-YBJ, and the hybrid ARGO-YBJ/WFCTA experiments in Tibet, all reported a knee-like structure below $1$ PeV for the spectrum of proton plus helium \citep{2013ICRC...33..338A,2015PhRvD..91k2017B,DAmone:2015wij,2015PhRvD..92i2005B}.
Specifically, the knee position of H+He flux is about $700$ TeV \citep{2015PhRvD..92i2005B}. Thus there is a tension for spectral measurements of the light component among different experiments. 

To pinpoint the knee position, multi-messenger observations such as ultra-high-energy diffuse $\gamma$-ray measurements, are expected to be very helpful. The Galactic diffuse $\gamma$-ray emission (DGE) is expected to be produced by interactions between CRs and the interstellar medium (ISM) and the interstellar radiation field (ISRF) during the propagation of CRs in the Milky Way. Recently, the DGE in the Galactic plane between 100 and 1000 TeV was for the first time measured by the Tibet-AS$\gamma$ experiment \citep{2021PhRvL.126n1101A}, which has attracted wide attention for possible physical origins \citep{2021ApJ...915...31K,2021arXiv210402838D,2021ApJ...919...93F,2021ApJ...914L...7L,2022FrPhy..1764501Q,Esmaili:2021yaw,2021PhRvD.104d3010K,Bouyahiaoui:2021rev,Carpet-3Group:2021ygp,Li:2021gah,2021Univ....7..141T, Maity:2021PRL}. 
This measurement also sheds new light on studying the individual nuclear spectrum of CRs.
In the analysis, the $\gamma$-ray events from 0.5 degrees of known point-like sources have been eliminated. Due to the fast cooling time of leptons \citep{1995PhRvD..52.3265A}, the propagation length of leptons is much shorter than nuclei. Therefore the ultra-high-energy DGE is thought to be predominantly generated from collisions of CR nuclei with the ISM \citep{1986A&A...157..223D,1997ApJ...478..225M,2010ApJ...722L..58S}. In order to produce the DGE spectrum measured by Tibet-AS$\gamma$ through hadronic interactions, it is required that Galactic CR sources accelerate nuclei to at least 10 PeV \citep{2021PhRvL.126n1101A}. 
This fact demonstrates that there are PeV accelerators (namely ``PeVatrons") in the Galactic disk, as revealed by recent ultra-high-energy $\gamma$-ray observations \citep{2021Natur.594...33C}.

{If an early knee as revealed by the Tibet experiment is correct, then the $\gamma$-ray observations suggest that an additional category of CR accelerators should exist in the Milky Way. This may have interesting implication on the sources of CRs. It is widely accepted that Galactic CRs can be largely accelerated in supernova remnants (SNRs) through the diffusive shock acceleration (DSA) process. However, soon after the proposal of DSA theory, it was realized that the maximum energy of accelerated particles in SNRs may be much less than PeV energies \citep{1983A&A...118..223L, 1983A&A...125..249L, 2021MNRAS.504.6096M}.
On the other hand, observations have shown that the $\gamma$-ray spectra of most SNRs fall off less than tens of TeV, which means that the maximum acceleration energy of CRs by SNRs is around $O(100)$ TeV. Other types of CR sources inside the Milky Way besides SNRs may be required.
Many possible candidates of PeVatrons have been proposed, including e.g., the supermassive black hole at the Galactic center, young massive clusters, mergers of neutron stars and so on. The HESS observations of  the diffuse emission in the Galactic center region showed no clear spectral cutoff up to 50 TeV \citep{2016Natur.531..476H}, indicating that the central supermassive black hole may be a PeVatron candidate. Massive star clusters such as the Cygnus cocoon are also potential PeVatron  \citep{2011Sci...334.1103A,2012ApJ...745L..22B,2018ApJ...861..134A,2021NatAs...5..465A}.
Particularly, the LHAASO observations of a 1.4 PeV photon from the direction of the Cygnus region further strengthen this argument \citep{2021Natur.594...33C}. Several other candidates of PeVatrons were also reported \citep{2021Natur.594...33C,HESS:2021zan}. 
It is likely that PeVatrons generally exist in the Galaxy. If so, the accelerated nuclei should also contribute to the locally observed CR spectra. The multiple source populations were also proposed by \citet{2013FrPhy...8..748G}.}

Based on the above discussion, we study the compatibility between diffuse $\gamma$-ray emission and the local measurements of light component spectra of CRs. Both the KASCADE and Tibet measurements will be discussed. The paper is organized as follows. Section 2 gives the model description. Section 3 presents the calculated results. We conclude our work in Section 4.

\section{Model Description}
\subsection{Propagation of CRs}
It has been recognized in recent years that the propagation of CRs in the Milky Way should depend on the spatial locations, as inferred by the HAWC observations of extended $\gamma$-ray halos around pulsars \citep{2017Sci...358..911A} and the spatial variations of the CR intensities and spectral indices from Fermi-LAT observations \citep{2016PhRvD..93l3007Y,2016ApJS..223...26A}. 
The spatially-dependent  propagation (SDP) model was also proposed to explain the observed hardening of CRs \citep{2012ApJ...752L..13T,2015PhRvD..92h1301T,2016PhRvD..94l3007F,2016ApJ...819...54G,2018ApJ...869..176L,2018PhRvD..97f3008G,2020ChPhC..44h5102T}, and also the large-scale anisotropies with the help of a nearby source \citep{2019JCAP...10..010L,2019JCAP...12..007Q,2020FrPhy..1624501Y}.

In the SDP model, the diffusive halo is divided into two parts, the inner halo (disk) and the outer halo. In the inner halo, the diffusion coefficient is much smaller than that in the outer halo, as indicated by the HAWC observations. The spatial diffusion coefficient $D_{xx}$ can be parameterized as
\begin{equation}
  D_{xx}(r,z, {\cal R} )= D_{0}F(r,z)\beta^{\eta} \left(\dfrac{\cal R}
  {{\cal R}_{0}} \right)^{\delta_{0}F(r,z)},
  \label{eq:diffusion}
\end{equation}
where
\begin{equation}
F(r,z) = {\dfrac{N_m}{1+f(r,z)}+\left[1-\dfrac{N_{m}}{1+f(r,z)}\right]}\left(\dfrac{{z}}{\xi{z}_{\rm h}} \right)^{n},  
\end{equation}
for ${{|z|} \leq \xi{z}_{\rm h}}$, and $F(r,z)=1$ otherwise. 
Here $r$ and $z$ are cylindrical coordinate, ${\cal R}$ is the particle's rigidity, $\beta$ is the particle's velocity in unit of light speed, $f(r,z)$ is the spatial distribution of the CR sources, 
$D_0$, $\delta_0$, $\eta$, and $N_m$ are constants.
%For the parameterization of $F(r,z)$ and $\delta(r,z)$, one can refer to \citep{2020ChPhC..44h5102T}.
The total half-thickness of the propagation halo is $z_h$, and the half-thickness of the inner halo is $\xi z_{h}$. The fitted parameters value for transport is listed in Table \ref{table_1}.

\begin{table*}
\centering
\caption{Parameters of the SDP propagation model of CRs.}
\begin{tabular}{ccccccc}
\hline
\hline
  ${D}_0~[\rm cm^{-2}s^{-1}]$ & $\delta_0$  & ${N}_{m}$  & $\xi$  & $n$ &  ${\cal \nu}_{\rm A}$~[km s$^{-1}$]   & ${z}_h$~[kpc] \\
\hline
  $4.87$ & $0.55$  &  $0.52$ & $0.1$ &  $4.0$  & $6$ & $5$ \\
\hline
\hline
\end{tabular}
\label{table_1}
\end{table*}

In this work, we adopt the diffusion re-acceleration model, with the diffusive re-acceleration coefficient $D_{pp}$, which correlated with $D_{xx}$ via $D_{pp}D_{xx} = \dfrac{4p^{2}v_{A}^{2}}{3\delta(4-\delta^{2}) (4-\delta)}$, where $v_A$ is the Alfv\'en velocity, $p$ is the momentum, and $\delta$ is the rigidity dependence slope of the diffusion coefficient \citep{1994ApJ...431..705S}. The numerical package DRAGON is used to solve the propagation equation of CRs \citep{2017JCAP...02..015E}. For energies smaller than tens of GeV, the fluxes of CRs are suppressed by the solar modulation effect. We use the force-field approximation \citep{1968ApJ...154.1011G} to account for the solar modulation.

\subsection{Background source distribution}
%As introduced in Section 1, the background source should include two categories as SNRs and PeVatrons. 
The spatial distribution of the background source distribution is assumed to be an axisymmetric form, which can be parameterized as
\begin{equation}
  f(r, z) = \left(\dfrac{r}{r_\odot} \right)^\alpha \exp \left[-\dfrac{\beta(r-r_\odot)}{r_\odot} \right] \exp \left(-\dfrac{|z|}{z_s} \right) ~,
\label{eq:radial_dis}
\end{equation}
where $r_\odot \equiv 8.5$ kpc represents the distance from the Galactic center to the solar system. Parameters $\alpha$ and $\beta$ are taken as $1.69$ and $3.33$ \citep{1996A&AS..120C.437C}. The density of the source distribution decreases exponentially along the vertical height from the Galactic plane, with $z_{s} = 200$ pc.

The injection spectrum of nuclei is assumed to be an exponentially cutoff power-law function of particle rigidity, $q({\cal R}) \propto {\cal R}^{-\nu} \exp(-{\cal R}/{{\cal R}_{\rm c}})$.

\subsection{Local source}
The fine structure of spectral hardening and softening at 200 GV and 14 TV respectively seems to be from a local source. The local source is also helpful to explain the evolution of large-scale anisotropies with energy \citep{2006APh....25..183E,2012JCAP...01..010B,2013ApJ...766....4P,2014ApJ...785..129K,2013APh....50...33S,2016PhRvL.117o1103A,2019JCAP...10..010L,2019JCAP...12..007Q}.
%Here the progenitor of Geminga, a SNRs, was introduced in this work. 
The injection process of the local source is approximated as a burst. The source injection rate as a function of time and rigidity is assumed to be
\begin{equation}
  Q({\cal R},t)=q_{0}\delta(t-t_0) \left(\dfrac{\cal R}{{\cal R}_0}
\right)^{-\gamma} \exp \left[-\dfrac{\cal R}{{\cal R}_{\rm c}}
\right] ~,
\label{eq:nearby}
\end{equation}
where ${\cal R}_{\rm c}$ is the cutoff rigidity and $t_0$ is the time of the supernova explosion. The propagated spectrum from the local source can thus be described by the Green's function as given in
\citet{1995PhRvD..52.3265A}.
%\begin{equation}
%  \varphi(\vec{r}, {\cal R}, t) = \int_{t_i}^{t} G(\vec{r}-\vec{r}^\prime, t-t^\prime, {\cal R}) Q_0(t^\prime) d t^\prime .
%\end{equation}
A Geminga-like source is found to well match the observations of both the energy spectra and anisotropies of CRs \citep{2019JCAP...10..010L,2022ApJ...926...41Z}. Assuming a distance of $0.3$ kpc and an age of $3\times10^5$ yr of the local source, the normalization is determined through fitting the CR energy spectra, which results in a total energy of $\sim 2.2\times 10^{50}$ erg for protons and $\sim 7.5\times 10^{49}$ erg for helium nuclei. If 10\% of kinetic energy is converted to accelerate CRs, the total energy of the source is estimated to be $\sim 3\times 10^{51}$ erg.

\section{Results}

Based on the above discussion, the fluxes of individual species in the solar neighborhood can be obtained from solving the propagation equation. Here two kinds of knee spectra of light component (protons and helium nuclei) are considered according to the measurements by KASCADE and Tibet experiments. 

\subsection{Spectra of individual species}

\begin{figure*}[!htb]
\centering
\includegraphics[width=0.48\textwidth]{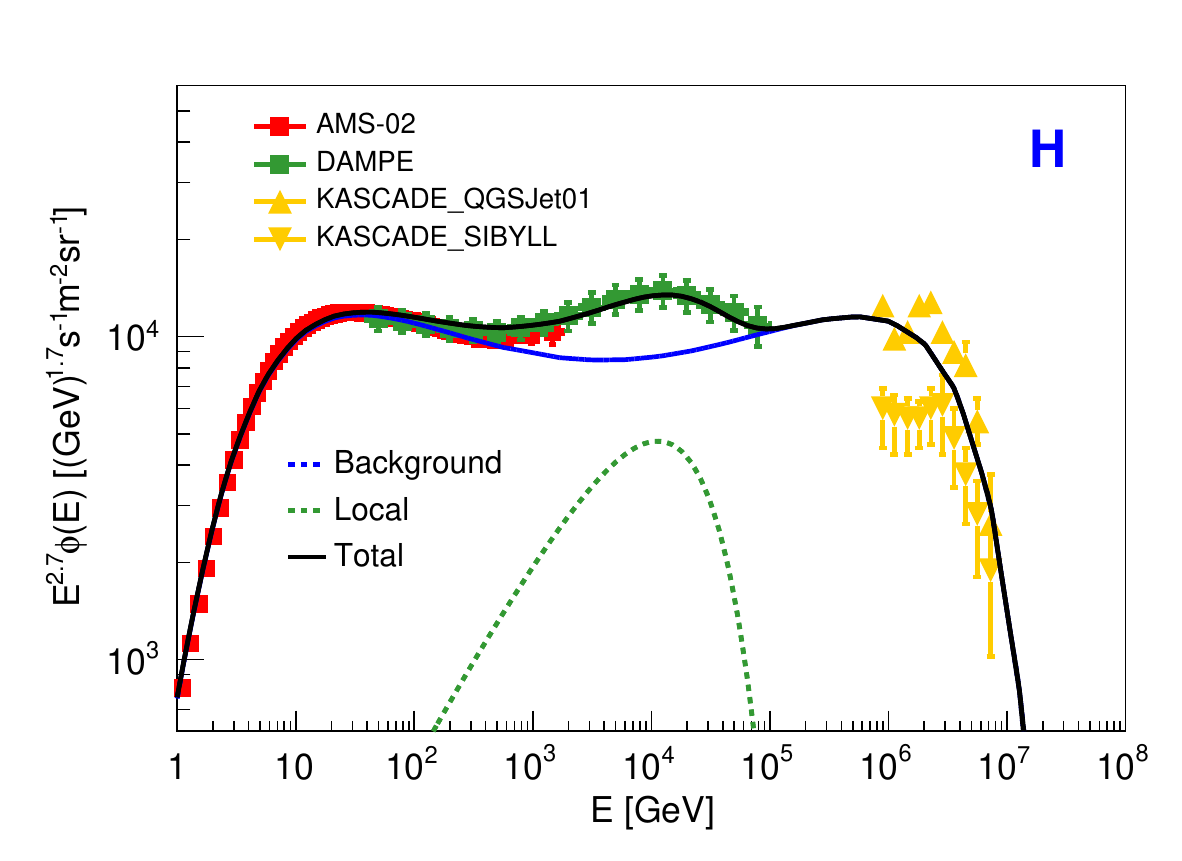}
\includegraphics[width=0.48\textwidth]{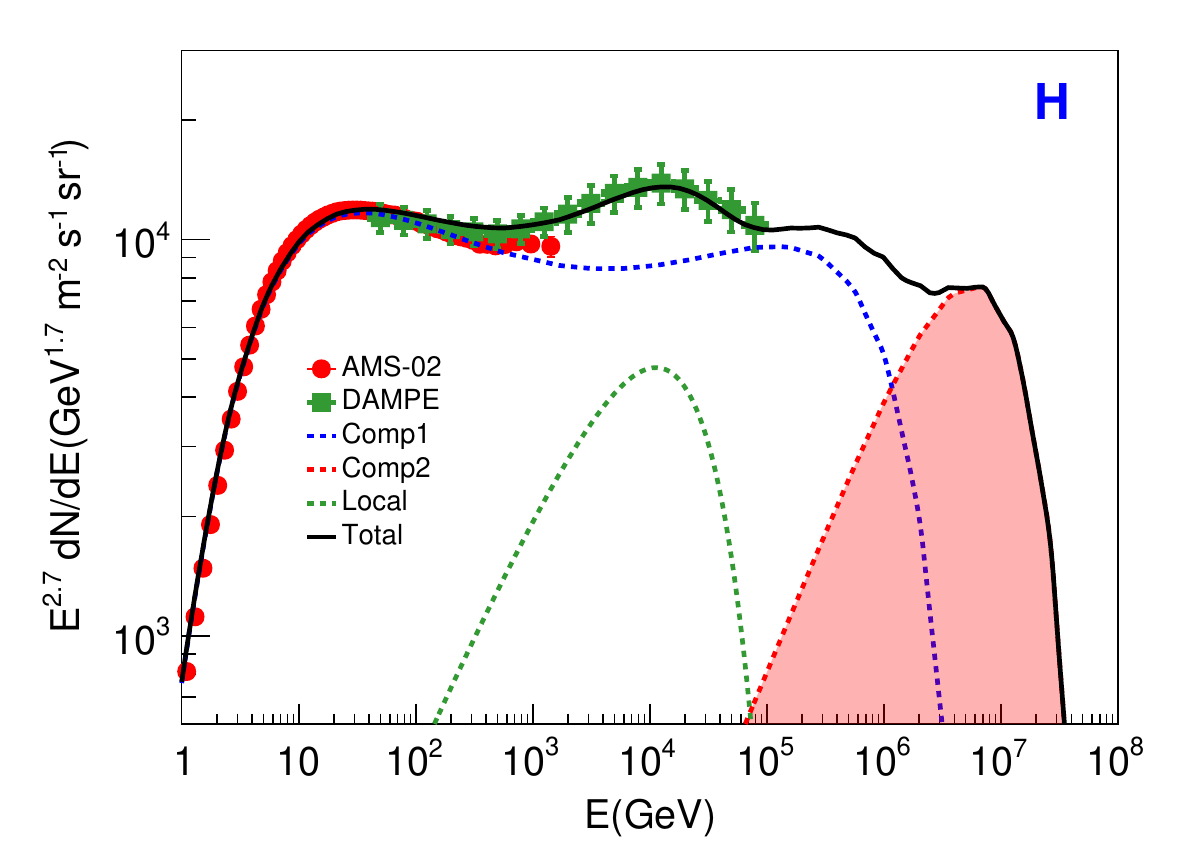}
\includegraphics[width=0.48\textwidth]{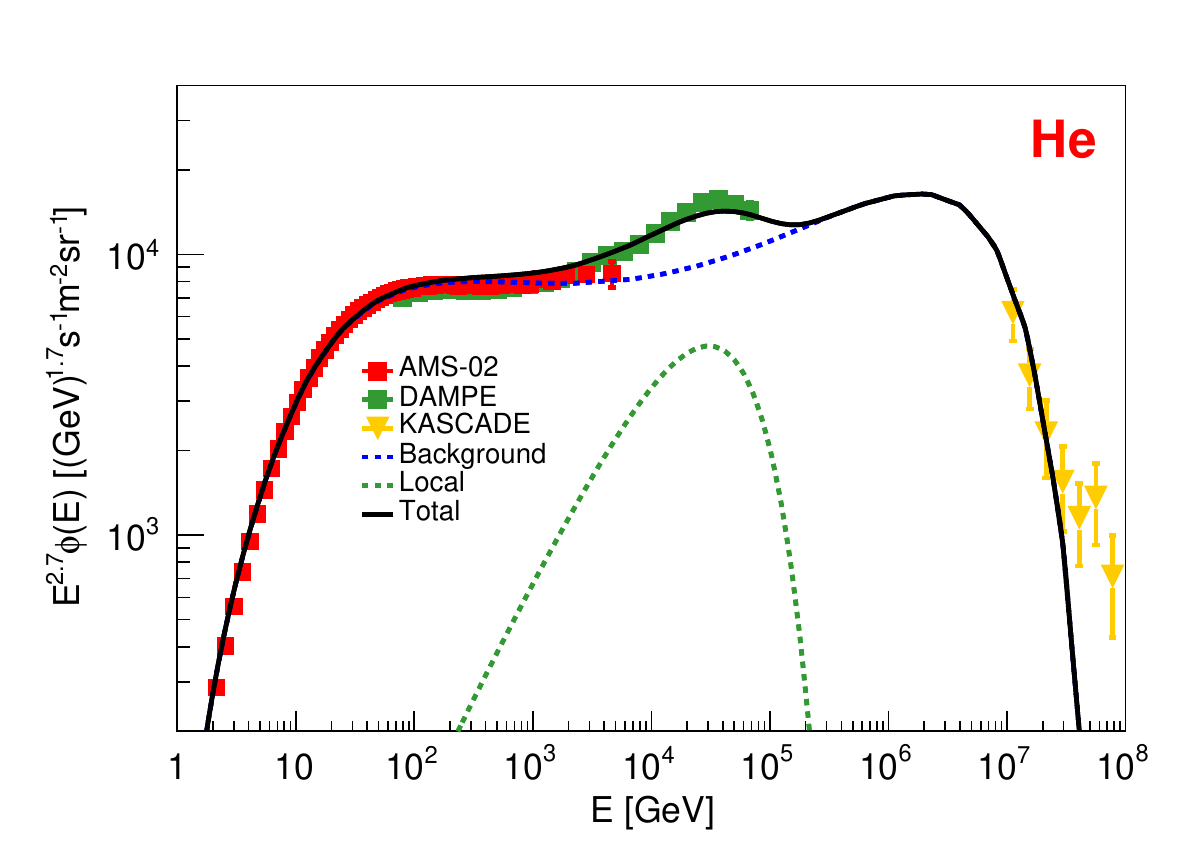}
\includegraphics[width=0.48\textwidth]{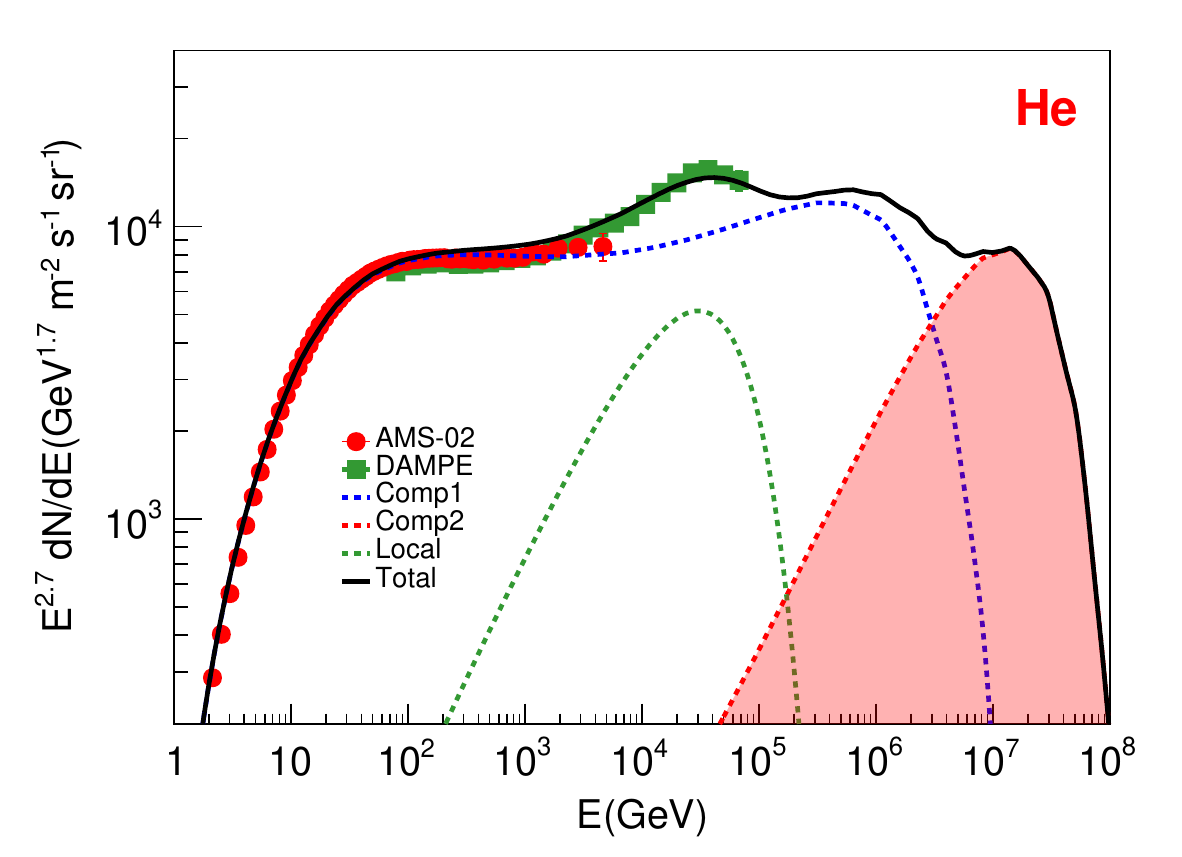}
\caption{The spectra of protons and helium nuclei for the cases corresponding to the KASCADE knee (left panels) and Tibet knee (right panels).
The data are from AMS-02 \citep{2015PhRvL.114q1103A,2017PhRvL.119y1101A}, DAMPE \citep{2019SciA....5.3793A,2021PhRvL.126t1102A}, and KASCADE \citep{2005APh....24....1A,2013APh....47...54A}.
  }
\label{fig:crs-spect}
\end{figure*}

Fig.~\ref{fig:crs-spect} shows the proton (top panels) and Helium (bottom panels) spectra from the model calculations, compared with the measurements by AMS-02 \citep{2015PhRvL.114q1103A,2017PhRvL.119y1101A},  DAMPE \citep{2019SciA....5.3793A,2021PhRvL.126t1102A}, and KASCADE \citep{2005APh....24....1A,2013APh....47...54A}. The left panels correspond to the fittings to the KASCADE measurements of protons and helium spectra, and the right panels correspond to the Tibet measurements on the H+He spectrum. In case of the Tibet knee, the contributions from the second source population (PeVatrons) are shown by red shaded regions. The hardenings and softenings of the spectra around several hundred GeV and $\sim14$ TeV are mainly due to the local source contribution. 
In Fig.~\ref{fig:crs-spect}, the blue dotted line represents the contributions from the background sources, with the proton cutoff rigidity of 4 PV and 980 TV for the KASCADE and Tibet results, respectively. The model parameters are listed in Table \ref{table-parm} and Table \ref{table-parm2}.
It can be seen that for the fitting to the KASCADE knee, the spectra of H and He in PeV energies are relatively smooth. For the fitting to the Tibet knee, potential spectral features (e.g., hardenings) exist due to the transition from the first source component (Comp1) to the second one (Comp2). {Although such features are not significant, they could be tested by the LHAASO experiment in the near future.}

\begin{table*}
\centering
\caption{Injection parameters of different components for fittings to the Tibet knee.}
\begin{tabular}{|c|c c c|c c c|c c c|}
\hline
  &  Comp1 & & & & local&  & Comp2 & &\\
\hline
  Element &  ${Q}_0~ [\rm m^{-2}sr^{-1}s^{-1}GeV^{-1}]^\dagger$ & $\nu$  & ${\cal R}_{\rm c}$ [TV]  & $q_0~ [\rm GeV^{-1}]$  
  & $\gamma$ &  ${\cal R}_{\rm c}$ ~[TV]   & ${Q}_0~ [\rm m^{-2}sr^{-1}s^{-1}GeV^{-1}]^\dagger$ & $\nu$  & ${\cal R}_{\rm c}$ [PV] \\
\hline
  P    & $4.42\times 10^{-2}$ & $2.40$   &  $980$ & $1.2\times 10^{52}$ &  $2.0$  & $18$ & $5.4\times 10^{-5}$ & $1.80$   &   $8$ \\
  He   & $2.72\times 10^{-3}$ & $2.31$  &   $980$ & $1.2\times 10^{51}$ &  $1.8$  & $18$ & $3.3\times 10^{-6}$ & $1.71$   &   $8$ \\
  C    & $5.65\times 10^{-5}$ & $2.20$  &   $980$ & $3.0\times 10^{50}$ &  $2.0$  & $18$ & $5.7\times 10^{-8}$ & $1.70$   &   $8$ \\
  N    & $1.36\times 10^{-5}$ & $2.37$  &   $980$ & $8.1\times 10^{49}$ &  $2.0$  & $18$ & $1.1\times 10^{-8}$ & $1.76$   &   $8$ \\
  O    & $8.51\times 10^{-5}$ & $2.27$  &   $980$ & $1.5\times 10^{50}$ &  $2.0$  & $18$ & $1.5\times 10^{-7}$ & $1.75$   &   $8$ \\
  Ne   & $1.05\times 10^{-5}$ & $2.20$  &   $980$ & $1.1\times 10^{50}$ &  $2.0$  & $18$ & $9.0\times 10^{-9}$ & $1.63$   &   $8$ \\
  Mg   & $1.91\times 10^{-5}$ & $2.26$  &   $980$ & $1.1\times 10^{50}$ &  $2.0$  & $18$ & $3.1\times 10^{-8}$ & $1.68$   &   $8$ \\
  Si   & $3.32\times 10^{-6}$ & $2.30$  &   $980$ & $1.1\times 10^{50}$ &  $2.0$  & $18$ & $3.0\times 10^{-9}$ & $1.74$   &   $8$ \\
  Fe   & $1.55\times 10^{-5}$ & $2.40$  &   $980$ & $1.8\times 10^{49}$ &  $2.0$  & $18$ & $1.9\times 10^{-8}$ & $1.65$   &   $8$ \\
\hline
%\hline
\end{tabular}
\label{table-parm}
\end{table*}

\begin{table*}
\centering
\caption{Injection parameters of different components for fittings to the KASCADE knee.}
\begin{tabular}{|c|c c c|c c c|}
%\hline
\hline
  &  \centering background & & & & local&\\
\hline
  Element &  ${Q}_0~ [\rm m^{-2}sr^{-1}s^{-1}GeV^{-1}]^\dagger$ & $\nu$  & ${\cal R}_{\rm c}$ [PV]  & $q_0~ [\rm GeV^{-1}]$  
  & $\gamma$ &  ${\cal R}_{\rm c}$ ~[TV]\\
\hline
  P    & $4.42\times 10^{-2}$ & $2.40$  &   $4$ & $1.2\times 10^{52}$ &  $2.0$  & $18$\\ 
  He   & $2.72\times 10^{-3}$ & $2.31$  &   $4$ & $1.2\times 10^{51}$ &  $1.8$  & $18$\\
  C    & $1.13\times 10^{-4}$ & $2.35$  &   $4$ & $1.0\times 10^{50}$ &  $2.0$  & $18$\\
  N    & $1.41\times 10^{-5}$ & $2.41$  &   $4$ & $4.0\times 10^{49}$ &  $2.0$  & $18$\\
  O    & $1.28\times 10^{-4}$ & $2.37$  &   $4$ & $1.5\times 10^{50}$ &  $2.0$  & $18$\\
  Ne   & $1.05\times 10^{-5}$ & $2.23$  &   $4$ & $3.0\times 10^{49}$ &  $2.0$  & $18$\\
  Mg   & $1.53\times 10^{-5}$ & $2.23$  &   $4$ & $1.5\times 10^{49}$ &  $2.0$  & $18$\\
  Si   & $3.02\times 10^{-6}$ & $2.34$  &   $4$ & $3.0\times 10^{49}$ &  $2.0$  & $18$\\
  Fe   & $1.40\times 10^{-5}$ & $2.28$  &   $4$ & $1.8\times 10^{49}$ &  $2.0$  & $18$\\
\hline
%\hline
\end{tabular}
\label{table-parm2}
\end{table*}

\subsection{H+He and all-particle spectra}

\begin{figure*}[!htb]
\centering
\includegraphics[width=0.48\textwidth]{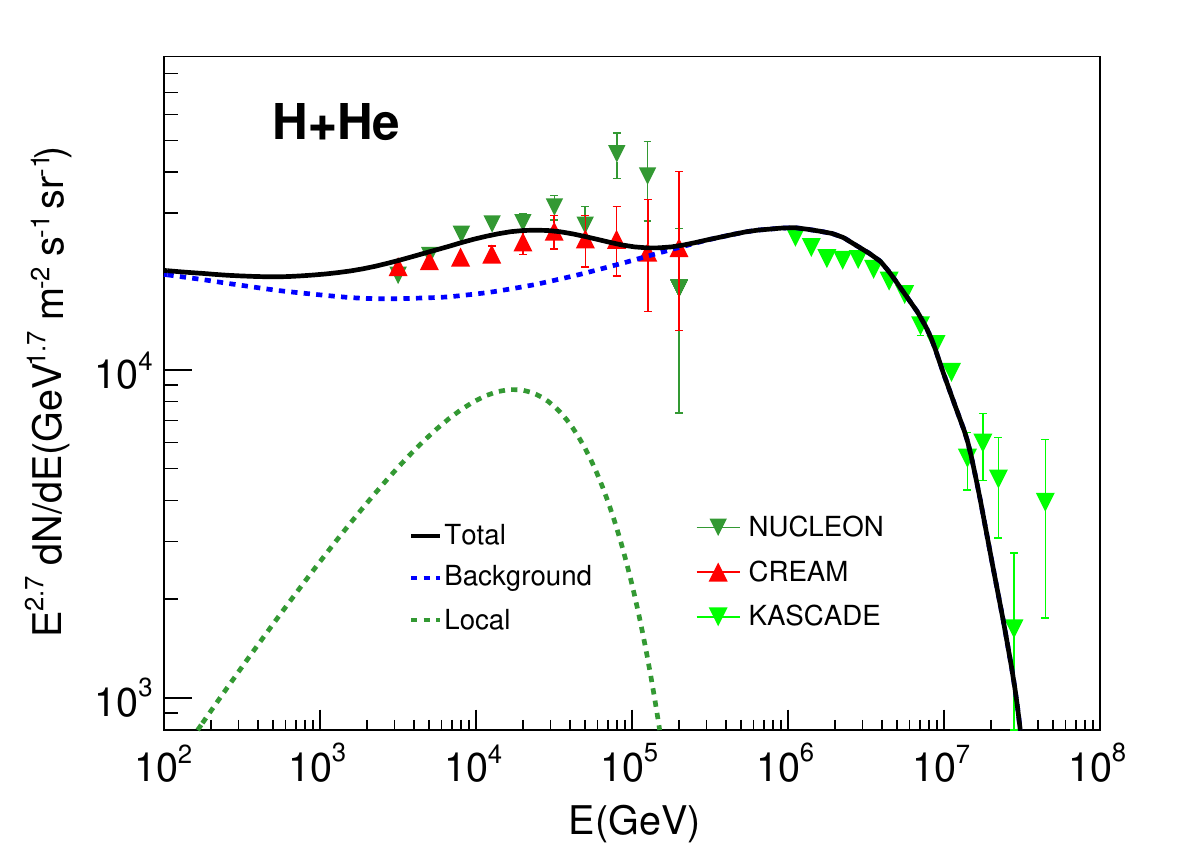}
\includegraphics[width=0.48\textwidth]{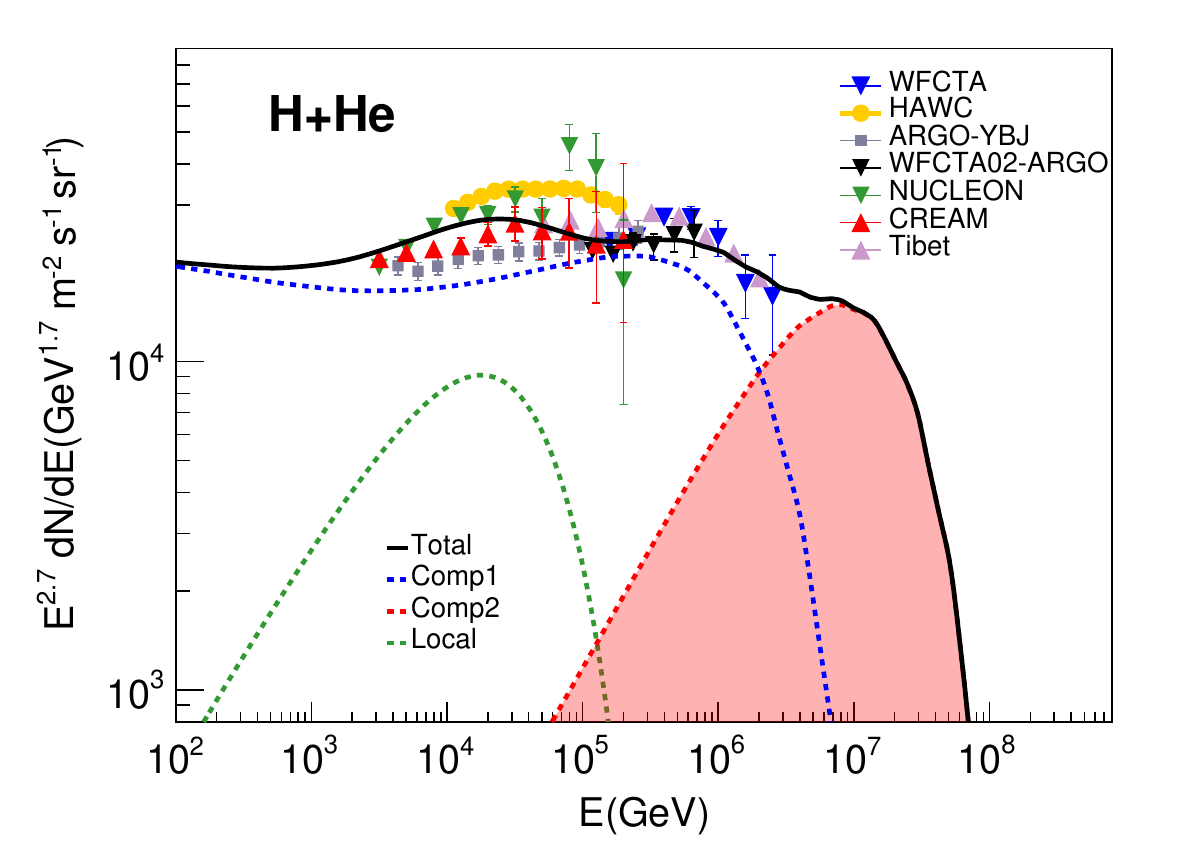}
\caption{The H+He spectra for the cases corresponding to the KASCADE knee (left panel) and Tibet knee (right panel). The measurements are from WFCTA \citep{2014ChPhC..38d5001B,2015PhRvD..92i2005B}, HAWC \citep{2019ICRC...36..176A}, ARGO-YBJ \citep{2015PhRvD..91k2017B}, NUCLEON \citep{2017JCAP...07..020A}, CREAM \citep{2017ApJ...839....5Y}, Tibet-AS$\gamma$ \citep{2013ICRC...33..338A}, and KASCADE \citep{2013APh....47...54A}.}
\label{fig:HHespect}
\end{figure*}

Fig.~\ref{fig:HHespect} shows the energy spectra of protons plus helium nuclei compared with measurements from direct detection \citep{2017ApJ...839....5Y,2017JCAP...07..020A} and indirect detection 
\citep{2014ChPhC..38d5001B,2015PhRvD..92i2005B,2019ICRC...36..176A,2015PhRvD..91k2017B,DAmone:2015wij,2013ICRC...33..338A} experiments. 
Similar to the individual species, the left panel shows the results corresponding to the KASCADE knee and the right one corresponds to the Tibet knee. 

\begin{figure*}[!htb]
\centering
\includegraphics[width=0.48\textwidth]{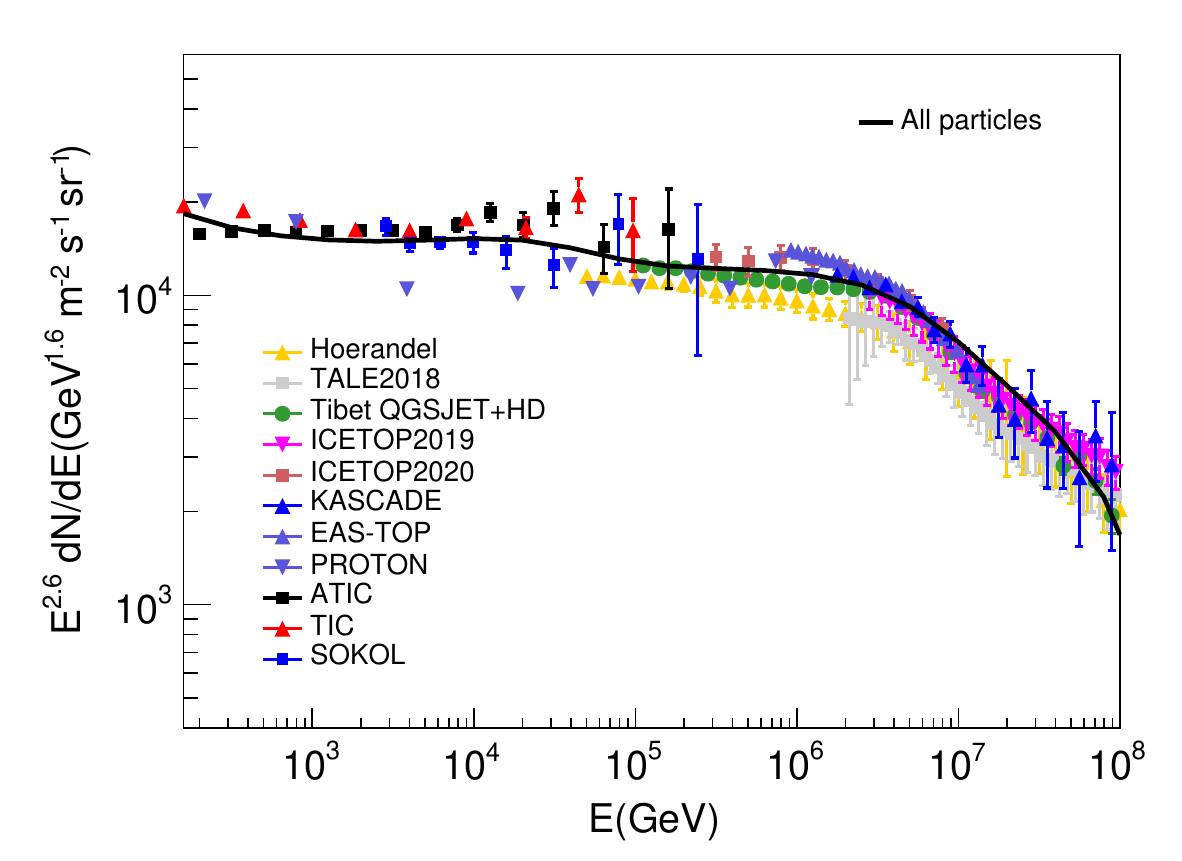}
\includegraphics[width=0.48\textwidth]{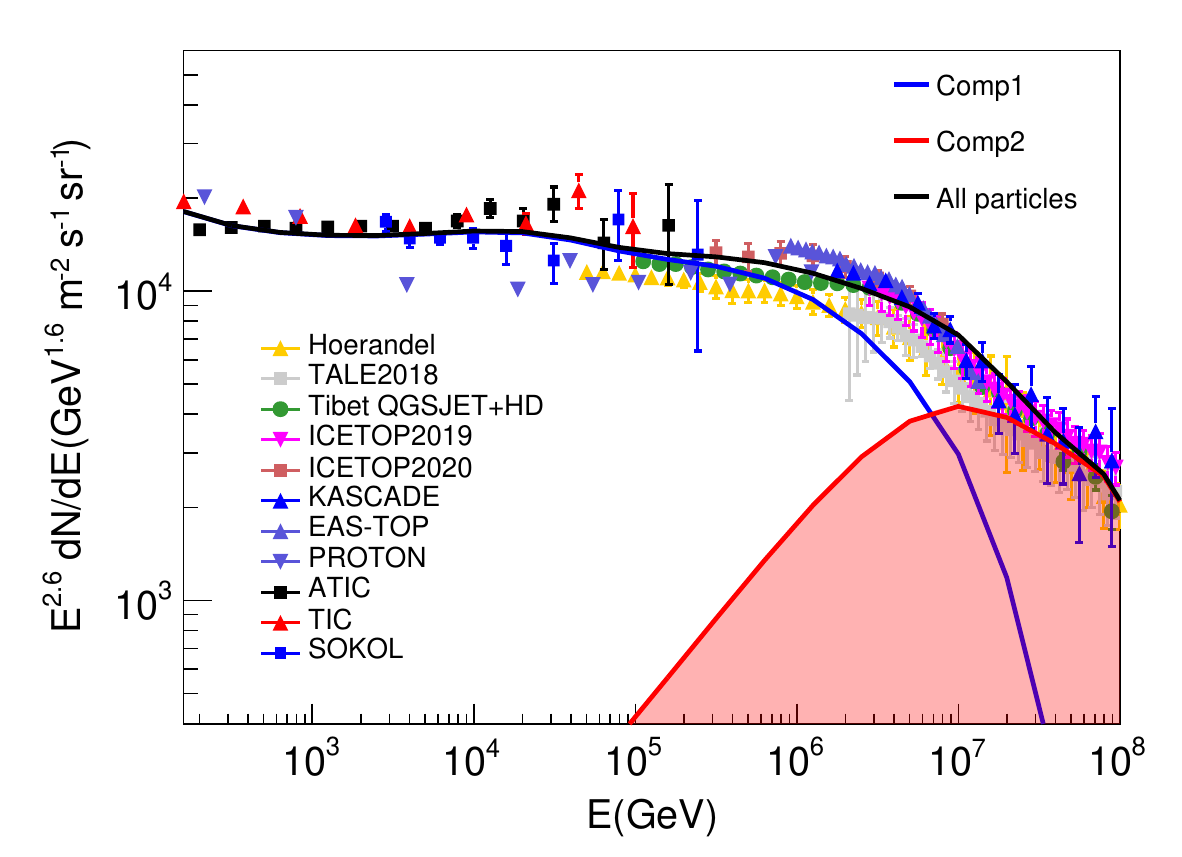}
\caption{The all-particle spectra for the cases corresponding to the KASCADE knee (left panel) and Tibet knee (right panel). The data are from Horandel \citep{2003APh....19..193H}, TALE \citep{2018ApJ...865...74A}, IceTop \citep{IceCube:2019hmk,IceCube:2020yct}, Tibet \citep{2008ApJ...678.1165A}, KASCADE \citep{2005APh....24....1A}, EAS-TOP \citep{1999APh....10....1E}, ATIC \citep{2009BRASP..73..564P}, SOKOL \citep{1993ICRC....1...17P}, TIC \citep{1997BRASP..61..922A} and PROTON \citep{1971ICRC....5.1746G}.}
\label{fig:all-spect}
\end{figure*}

Fig.~\ref{fig:all-spect} shows the model calculation of the all-particle spectrum, compared with data \citep{2003APh....19..193H,2018ApJ...865...74A,IceCube:2019hmk,IceCube:2020yct,2008ApJ...678.1165A,2005APh....24....1A,1999APh....10....1E,2009BRASP..73..564P,1993ICRC....1...17P,1997BRASP..61..922A,1971ICRC....5.1746G}. As shown in the left panel, a 4 PV knee consistent with KASCADE can also well describe the all-particle spectrum up to several times of $10^7$ GeV. If the knee is as early as $\sim 980$ TV, the all-particle spectrum deviates clearly from the data. After adding the contribution from Comp2, the all-particle spectrum can be well reproduced. 
  
\begin{figure*}[!htb]
\centering
\includegraphics[width=0.48\textwidth]{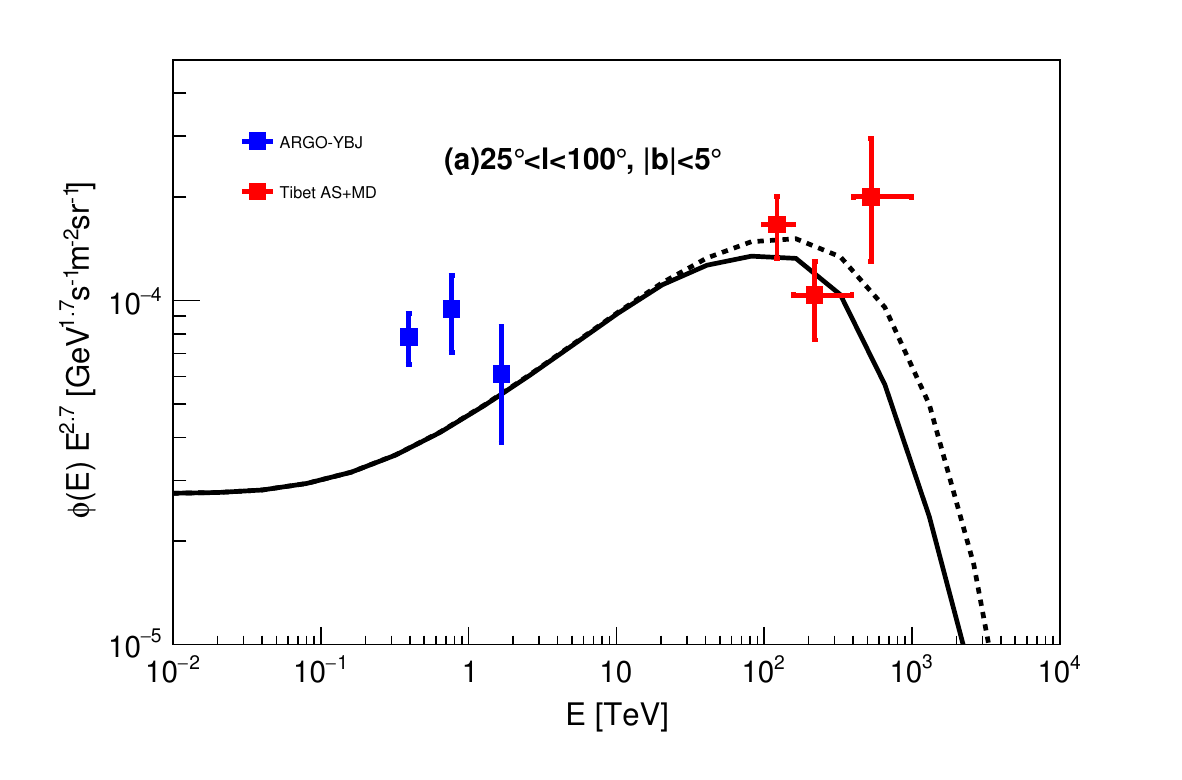}
\includegraphics[width=0.49\textwidth]{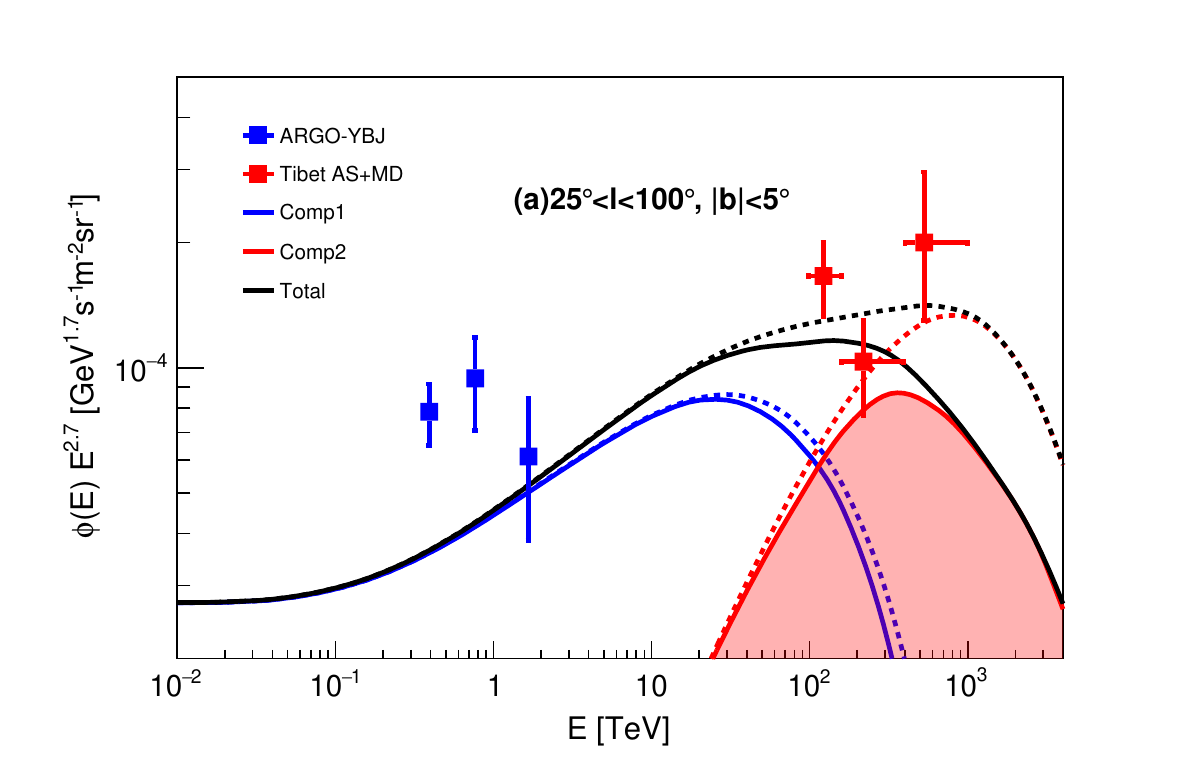}
\includegraphics[width=0.48\textwidth]{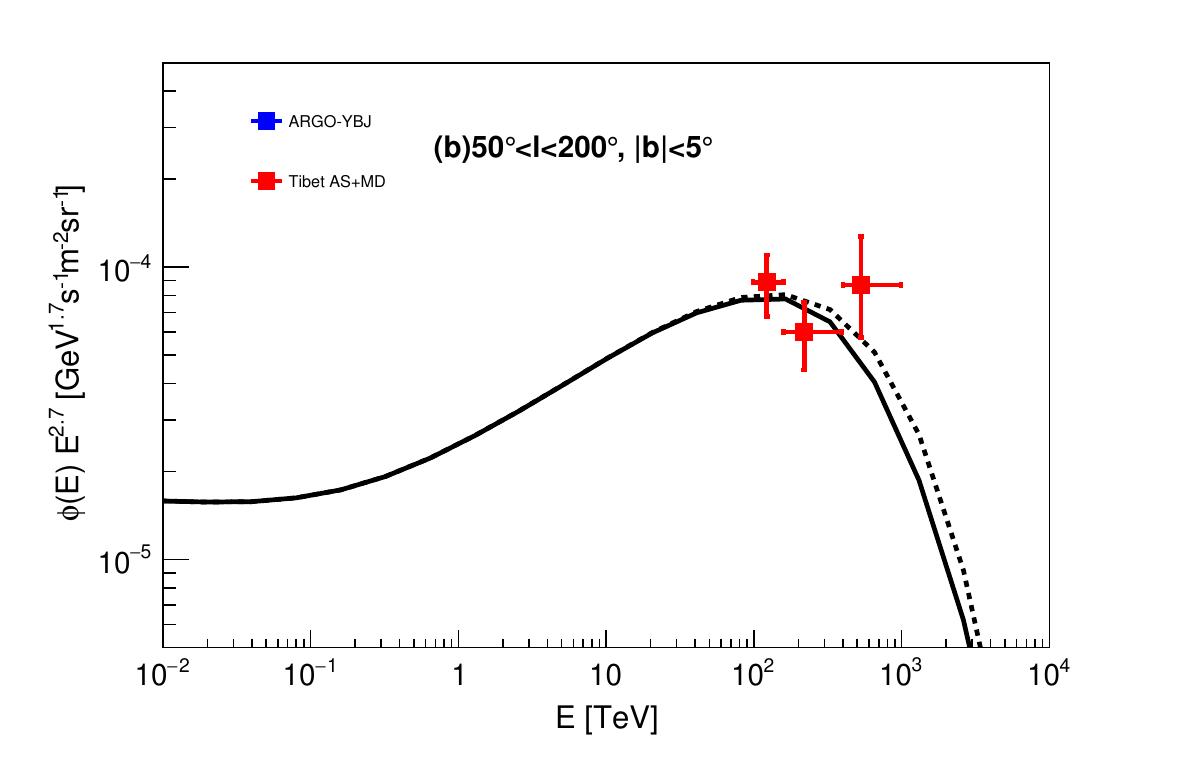}
\includegraphics[width=0.49\textwidth]{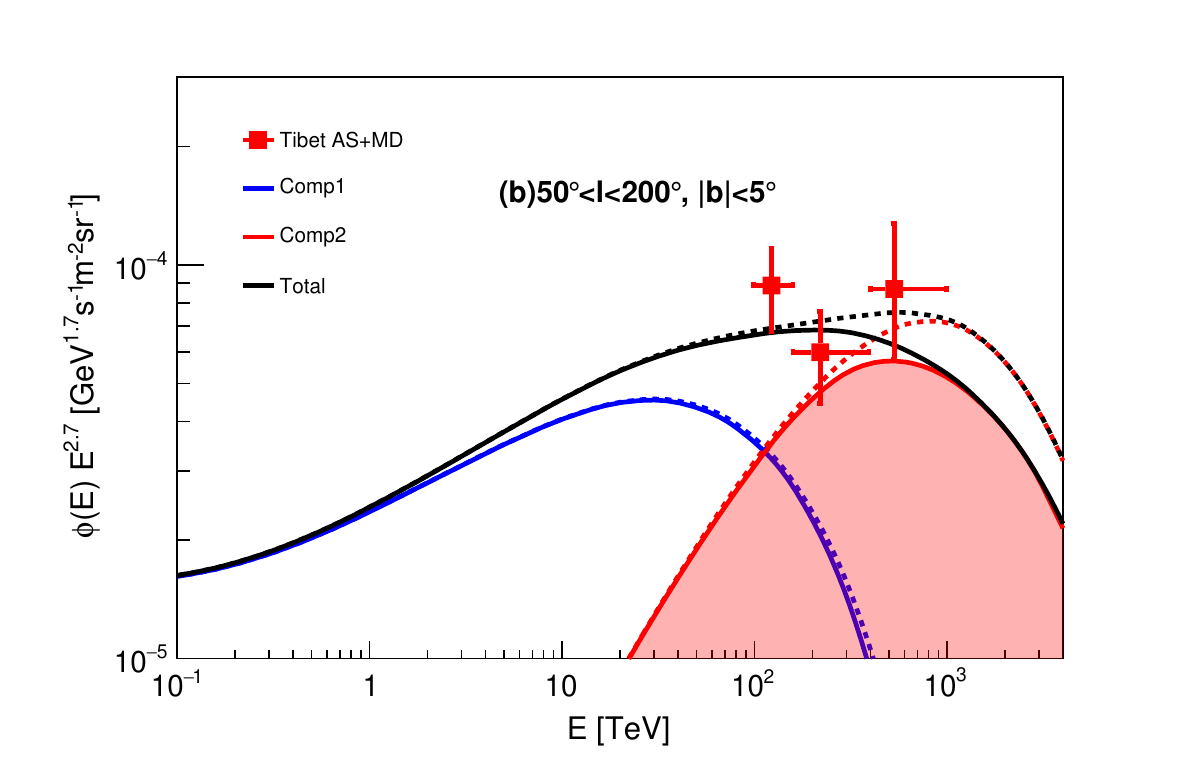}
\caption{Diffuse $\gamma$-ray spectra from the model calculation, compared
with the measurements by ARGO-YBJ \citep{2015ApJ...806...20B} and Tibet
AS$\gamma$ \citep{2021PhRvL.126n1101A}. Left panels correspond to the KASCADE 
knee, and right panels correspond to the Tibet knee.}
\label{fig:gamma-spect}
\end{figure*}

\subsection{Diffuse $\gamma$-rays}

{The Galactic DGE is assumed to be produced by interactions between CRs and the ISM as well as the ISRF, during the propagation of CRs in the Milky Way. The ultra-high-energy DGE in the Galactic plane at the energy of 957 TeV was for the first time measured by the Tibet-AS$\gamma$ experiment \citep{2021PhRvL.126n1101A}. While it is possible that faint unresolved sources may contribute to part of the measured emission \citep[e.g.,][]{2022ApJ...928...19V}, we do not consider this part in our model calculation.}
%similar to other works \citep{2022A&A...661A..72B,2022FrPhy..1764501Q,2022arXiv220315759D}. In addition, the measurements may be affected by the uniform exposure and zenith angle of sky. For sake of simplicity, we give the ideal theoretical results and omit the effect of experimental uncertainty.
In Fig.~\ref{fig:gamma-spect}, we show the results in two different sky regions as defined in \citet{2021PhRvL.126n1101A}: (a) the Inner Galactic Plane with $25^\circ<l<100^\circ,~|b|<5^\circ$, 
and (b) the Outer Galactic Plane with $50^\circ<l<200^\circ,~|b|<5^\circ$. 
At ultra-high energies ($E\gtrsim100$ TeV), the absorption of $\gamma$-rays due to pair production with ISRF becomes important \citep{2006A&A...449..641Z}, which results in reduction of the DGE flux, as can be seen by the dashed (without absorption) and solid (with absorption) lines. 
The left panels correspond to the KSACADE knee and the right panels correspond to the Tibet knee. In both cases, the diffuse $\gamma$-ray fluxes can be properly reproduced. 

\section{Discussion and Conclusion}

The knee for individual species have been observed below PeV and around 4 PeV by Tibet and KASCADE experiments. The diffuse $\gamma$-rays can be a useful probe to test the ambiguous CR measurements. In this work, we simultaneously calculated the nuclei and diffuse $\gamma$-rays, assuming different knee features of light components as measured by KASCADE and Tibet experiments. When adopting the KASCADE knee, consistent results with the all-particle spectrum and the DGE can be obtained. On the other hand, if we adopt the Tibet knee of the light components in the model, an additional CR source component is clearly required to explain the all-particle spectrum and the DGE. The additional source population might be the generally existing PeVatrons as revealed by recent $\gamma$-ray observations \citep{2021Natur.594...33C}. 
These two scenarios show differences in the spectra of individual species in PeV energy ranges and DGE, which may be distinguished by future observations. {Note that constrained by the sharp break of H+He spectrum at $\sim700$ TeV as indicated by the Tibet experiments and the knee structure at $\sim 4$ PeV for the all particle spectrum, the spectral index of ``Comp2'' is required to be hard (1.8 as given in Table \ref{table-parm2}). Even for a very hard spectrum, we find that the spectra due to the sum of the two components show smooth transitions. Some kind of fine tuning of model parameters between these two components is necessary. Nevertheless, precise measurements of the spectra to an accuracy of 10\% level may crucially test whether there are spectral features of individual CRs in PeV energy range.}

The existence of new structures is very important to decipher the origin of CRs. It means that there should be several groups of accelerators \citep{2013FrPhy...8..748G}. Given more and more evidence of the existence of various types of high-energy sources, complicated spectral structures are expected to be natural. More precise measurements of the spectra of individual species, diffuse $\gamma$-rays, and perhaps diffuse neutrinos can further test the origin of CRs in future.

\section*{Acknowledgements}
This work is supported by the National Key Research and Development Program of China
(Nos. 2018YFA0404203) and the National Natural Science Foundation of China (Nos. 12275279, 12220101003).

\bibliographystyle{aasjournal}
\bibliography{refs}

\end{document}